# Neighborhood Overlap-Aware High-Order Graph Neural Network for Dynamic Graph Learning

Ling Wang
*School of Computer Science and Technology*
*Chongqing University of Posts and Telecommunications*
Chongqing, China
wangling1820@126.com

*Abstract*—Dynamic graph learning (DGL) aims to learn informative and temporally-evolving node embeddings to support downstream tasks such as link prediction. A fundamental challenge in DGL lies in effectively modeling both the temporal dynamics and structural dependencies of evolving graph topologies. Recent advances in Dynamic Graph Neural Networks (DGNNs) have obtained remarkable success by leveraging message-passing mechanisms to capture pairwise node interactions. However, these approaches often overlook more complex structural patterns, particularly neighborhood overlap, which can play a critical role in characterizing node interactions. To overcome this limitation, we introduce the Neighborhood Overlap-Aware High-Order Graph Neural Network (NO-HGNN), which is built upon two key innovations: (a) computing a correlation score based on the extent of neighborhood overlap to better capture complex node interactions; and (b) embedding this correlation directly into the message-passing process of high-order graph neural networks in the DGL. Experiments on two real-world dynamic graphs show that NO-HGNN achieves notable improvements in link prediction accuracy, outperforming several state-of-the-art approaches.

I. INTRODUCTION

DYNAMIC graphs (DGs) are widely encountered in real-world contexts [1-5], such as social networks [6-10], e-commerce systems [11-17], and biological networks [18-24]. Dynamic graph learning (DGL) [25-28] plays a crucial role in numerous applications, including user behavior prediction [29-30] and protein interactions analysis [31-33]. The primary challenges in building an effective DGL framework lie in devising efficient message passing to model the complex inter-node dependencies, such as neighborhood overlap correlations. This message passing mechanism in the DGL framework ensures that nodes with close relationships are represented by similar embedding vectors, thereby enhancing the quality of DGL and the performance of the downstream tasks like link prediction. improving both the expressiveness of learned representations and the overall performance on link prediction.

Dynamic Graph Neural Networks (DGNNs) have demonstrated strong performance in DGL [34-36]. A core component across various DGNNs is message passing, which updates node representations by aggregating information from neighboring nodes [37, 38]. For example, models like WDGCN [39] and EvolveGCN [40] use spectral-based filters, where each node refines its representation by aggregating features from its neighboring nodes. However, these methods typically rely on uniform or average weighting during message passing, which limits their ability to fully exploit the rich semantic information in node features. To address this, more recent approaches [41-45] enhance message-passing by considering feature correlations between nodes. For instance, DySAT [46] learns feature correlation scores to assign higher weights to more relevant neighbors, improving the quality of the aggregated representation. Despite these advancements, current message passing mechanisms in DGNNs still fail to consider the neighborhood overlay.

To address the aforementioned challenges, this study proposes a Neighborhood Overlap-Aware High-Order Graph Neural Network (NO-HGNN) for dynamic graph learning. NO-HGNN is designed to capture overlap correlations by leveraging structural information, which is seamlessly integrated into the message passing framework. Additionally, it utilizes high-order graph neural networks (HGNN) [47–51] to enhance the message propagation process and generate expressive dynamic graph (DG) representations.

The key contributions of this work are summarized:
a) *Neighborhood overlaps are captured*. Structural features are extracted to estimate the overlap correlation score by modeling the neighborhood overlap within DG, and
b) *Message passing is guided*. The overlap correlation score is integrated into the message passing process of the HGNN to enhance the learning of DG representations.

The remainder of the paper is organized as: Section II, we present the problem definition, introduce the tensor product, and review the basics of HGNN. Section III introduces the NO-HGNN model in detail. The experimental setup and corresponding results are presented and discussed in Section IV. Finally, Section V concludes the paper by summarizing our contributions and outlining possible future work.

## II. PRELIMINARIES

This section begins by outlining the problem defination addressed through DG representation learning, followed by an introduction to the necessary HGNN's preliminaries.

### A. Problem Defination

In this study, a DG is defined as $G = \{G_t\}_{t=1}^{T}$, where each snapshot $G_t = (V, E_t, X_t)$ represents the graph at the $t$-th time step, $t \in [1, T]$. Here, $V$ represents the node set with size $|V|=N$, $E_t$ denotes the edge set at time slot $t$, and $X_t$ contains the feature vectors of all nodes at that time. For each graph snapshot $G_t$, the edge set $E_t$ is encoded as an adjacency matrix $A_t \in \mathbb{R}^{N \times N}$. Accordingly, the entire dynamic graph can be characterized by a feature tensor $\mathbf{X}=[x_{ijt}] \in \mathbb{R}^{N \times F \times T}$ and an adjacency tensor $\mathbf{A}=[a_{ijt}] \in \mathbb{R}^{N \times N \times T}$. This study aims to learn a node embedding tensor $\mathbf{H}=[h_{it}] \in \mathbb{R}^{N \times F \times T}$ that effectively captures spatiotemporal patterns, which is then used to predict potential links between nodes over time.

### B. High-order Graph Neural Network

The High-order Graph Neural Network (HGNN), constructed based on the tensor product, offers a straightforward yet effective approach for modeling dynamic graphs (DGs). In our proposed framework, the DG is naturally represented in tensor form, and HGNN serves as the backbone model for capturing dynamic patterns and learning node representations for the link prediction task. The following section provides a detailed introduction to the tensor product and the HGNN.

**Tensor product.** Let $\mathbf{X} \in \mathbb{R}^{I \times J \times T}$ and $\mathbf{Y} \in \mathbb{R}^{J \times K \times T}$ be two third-order tensors, and let $M \in \mathbb{R}^{T \times T}$ be a matrix. The tensor product, denoted by $\mathbf{X} \circledast \mathbf{Y} \in \mathbb{R}^{I \times K \times T}$, is defined as:

$$\mathbf{X} \circledast \mathbf{Y} = \left( \left( \mathbf{X} \times_3 M \right) \otimes \left( \mathbf{Y} \times_3 M \right) \right) \times_3 M^{-1}, \tag{1}$$

where $\times_3$ is the tensor mode-3 product, which facilitates information mixing along the third-order dimension, and $\otimes$ denotes the tensor face-wise product as defined in [52].

**HGNN.** As introduced in [52], the High-order Graph Neural Network (HGNN) leveraging the tensor product is defined as follows:

$$\begin{cases} \mathbf{H}^{(0)} = \mathbf{X}, \\ \mathbf{H}^{(l)} = \sigma \left( \mathbf{P} \circledast \mathbf{H}^{(l-1)} \circledast \mathbf{W}^{(l)} \right), \end{cases} \tag{2}$$

where $\sigma$ is a non-linear activation function, the node feature tensor is denoted as $\mathbf{X}$, while $\mathbf{W}^{(l)} \in \mathbb{R}^{F \times F \times T}$ is a trainable weight tensor at the $l$-th HGNN layer, and $\mathbf{H}^{(l)} \in \mathbb{R}^{N \times F \times T}$ is the node embedding tensor. $\mathbf{P}$ is the adjacent tensor responsible for message passing, which encodes the structural dependencies.

## III. PROPOSED MODEL

In this paper, we propose a Neighborhood Overlap-Aware High-Order Graph Neural Network (NO-HGNN) for link prediction on dynamic graphs. The proposed model is designed to address the limitations of existing message-passing methods by explicitly incorporating neighborhood overlap information into the aggregation process. NO-HGNN consists of two key components: a) Structural Feature Learning Module, which encodes edge information into node-level structural feature tensors, capturing rich local structural patterns across time; b) Neighborhood Overlap-Aware Module, which computes neighborhood overlap correlations based on structural features. Furthermore, the model leverages high-order graph neural networks (HGNNs) built upon tensor operations to perform expressive message passing, enabling dynamic graph learning. The detailed design of each module is presented in the following sections.

### A. Structural Feature Learning Module

In this study, we propose a structural feature learning module aimed at extracting structural features by modeling neighborhood overlap. Previous studies [53-57] have shown that one effective approach for capturing such structural information is through the exponentiation of the adjacency tensor $\mathbf{A}$, which reflects higher-order connectivity patterns within the graph. Building upon this idea, our method further enhances the representation capacity by leveraging tensor-based operations to capture interaction among nodes. Specifically, to count the neighborhood overlap directly, the unnormalized tensor $\mathbf{A}$ is employed in the procedure to account for the multi-hop neighborhood overlap as follows:

$$\mathbf{B} = \sum_{k=1}^{K} \mathbf{A}^k. \tag{3}$$

For clarity, the tensor form of (3) can be presented in matrix form, in which each front slice $Z_t$ in $\mathbf{Z}$ is obtained as:

$$B_t = \sum_{k=1}^{K} A_t^k, \tag{4}$$

where $K$ decides the maximum number of hops, and $A^k$ is responsible for capturing the $k$-hop neighborhood overlap.

While this method effectively aggregates multi-hop neighborhood information, it is inherently limited by its reliance on predefined edge structures, making it less effective in capturing intricate or latent structural dependencies. To address this issue,

NO-HGNN incorporates a structural feature generator, which aims to learn more expressive and generalized structural representations. Concretely, a learnable generator $g_\theta$ is applied to each node, computing its structural features directly from the adjacency tensor $\mathbf{B}=[b_{ijt}]\in\mathbb{R}^{N\times N\times T}$ of DG as:

$$o_{it} = g_\theta\left(\sum_{j\in\mathcal{N}(i)} g_{edge}(b_{ijt})\right), \quad (5)$$

where $o_{it}$ denotes the structural feature of $i$-th node at $t$-th time slot, and $g_\theta$ is a learnable function composed of two multi-layer perceptrons (MLPs), $g_{edge}$ is responsible for extracting edge-level features. Through this design, NO-HGNN requires only the adjacency tensor $\mathbf{B}$ as input and is capable of adaptively learning a structural feature tensor $\mathbf{O}=[o_{it}]\in\mathbb{R}^{N\times F\times T}$ after considering the neighborhood overlap.

*B. Neighborhood Overlap Awareness Module*

The Neighborhood Overlap Awareness (NOA) module is designed to estimate the relational strength between node pairs by measuring their neighborhood overlap and to incorporate this information into the message passing in HGCNN. To evaluate the impact of neighborhood overlap during message propagation, the overlap score between nodes $i$ and $j$ is computed as:

$$\hat{p}_{ijt} = o_{it} o_{jt}^\top, \quad (6)$$

where $\hat{p}_{ijt}$ represents the degree of overlap between the neighborhoods of nodes. By computing these scores across all node pairs and all time steps, we construct a score tensor $\hat{\mathbf{P}}$ that captures the neighborhood overlap patterns.

Traditional DGNNs typically compute aggregation weights based only on directly connected node pairs, overlooking the indirect influence introduced by overlapping neighborhoods. For this limitation, we introduce a learnable scheme to model neighborhood overlap interventions. Based on the initial correlation scores $\hat{p}_{ijt}$, we further compute normalized correlation weights as follows:

$$p_{ijt} = \phi(\hat{p}_{ijt}) = \frac{\exp(\hat{p}_{ijt})}{\sum_{k\in N_t(i)}\exp(\hat{p}_{ikt})}, \quad (7)$$

where $\phi$ denotes the SoftMax function, which is applied to normalize the overlap scores, and $p_{ijt}$ represents the neighborhood overlap correlation between nodes. Through this normalization process, we obtain a refined aggregation weight tensor $\mathbf{P}$, which captures the relative importance of neighborhood overlap across node pairs. Then, the tensor $\mathbf{P}$ is incorporated into the high-order graph neural network (HGNN) to facilitate message passing, enabling node embedding learning in DG to obtain the embedding tensor $\mathbf{H}$.

*C. Model Optimization*

In this study, we learn dynamic graph representations. i.e., $\mathbf{H}$, tailored for the link prediction task. For a potential link between the $i$-th nodes and $j$-th node at time slot $t$, their corresponding node embedding vectors are first concatenated and then passed through a multi-layer perceptron (MLP) to estimate the link probability, as follows:

$$\hat{y}_{ijt} = \varphi(h_{it} \| h_{jt}), \quad (8)$$

where $\hat{y}_{ijt}$ represents the predicted link probability for the concatenation of the two node embedding vectors. The function $\varphi$ is an MLP used to compute the link probability, and $\cdot\|\cdot$ denotes the concatenation operation applied to the node embedding vectors. To optimize the model parameters, a learning objective is defined:

$$L = \frac{1}{|\Lambda|}\sum_{y_{ijt}\in\Lambda}\left(y_{ijt}\log\hat{y}_{ijt} + (1-y_{ijt})\log(1-\hat{y}_{ijt})\right) + \beta\|\Omega\|_2, \quad (9)$$

where $L$ denotes the overall loss. $y_{ijt}$ is the real link indicating the existence of a link between the $i$-th node and $j$-th node at the $t$-th time slot. $\Omega$ represents the set of model parameters to be optimized, and $\Lambda$ denotes the training set containing labeled node pairs. $\|\cdot\|_2$ corresponds to the $L_2$ regularization applied to the parameters, with a regularization coefficient $\beta$.

## IV. EXPERIMENTS

In this section, we introduce two dynamic graph datasets used to evaluate the effectiveness of the proposed NO-HGNN model and benchmark its performance against six state-of-the-art baselines.

*A. Experiment Settings*

**Datasets.** This study utilizes three DG datasets, as summarized in Table I. Each dataset is partitioned into training, validation, and test sets following a 70%-20%-10% split. Consistent with the approach in [58], negative sampling is applied to construct the training data for link prediction.

TABLE I. Experimental Data Statistics

| No. | Dataset | Nodes | Edges | Time slots |
|---|---|---|---|---|
| D1 | ask-ubuntu | 3,748 | 159,817 | 73 |
| D2 | bitcoin-alpha | 3,783 | 24,187 | 32 |

TABLE II. Comparison results on F1-Scores and Accuracy. The best results are in **bold**, and the suboptimal results are underlined for clarity.

| Model | D1 | | D2 | |
|---|---|---|---|---|
| | F1-Score | Accuracy | F1-Score | Accuracy |
| M1 | 0.7641 | 0.7300 | 0.7220 | 0.6962 |
| M2 | 0.7519 | 0.7173 | 0.7300 | 0.7039 |
| M3 | 0.7913 | 0.7676 | 0.7844 | 0.7553 |
| M4 | 0.8069 | 0.7784 | 0.7901 | 0.7695 |
| M5 | 0.8166 | 0.7983 | 0.8093 | 0.7846 |
| M6 | <u>0.8247</u> | <u>0.8002</u> | <u>0.8134</u> | <u>0.7935</u> |
| M7 | **0.8301** | **0.8125** | **0.8266** | **0.8094** |

*Evaluation Protocols.* In this study, the performance of the proposed model on the link prediction task in dynamic graph representation learning is evaluated using F1-score [59-65] and Accuracy [65-68] as the primary evaluation metrics.

*Baseline Models.* To assess the effectiveness of the proposed NO-HGNN, comparisons are conducted with six representative DG learning models. which include two static graph neural networks (i.e., GCN (M1), LightGCN (M2)), four dynamic graph neural networks (i.e., WDGCN (M3)), EvolveGCN (M4), DySAT (M5), FSTGCN (M6)). The proposed model, NO-HGNN, is referred to as M7.

*Experimental Setup.* To guarantee fairness in model comparison, we adopt the following training configuration:
a) The proposed NO-HGNN model is implemented using PyTorch and trained on four NVIDIA 2080Ti GPUs;
b) All baseline and comparison models are optimized using the Adam optimizer. The learning rate is selected from the set {0.1, 0.01, 0.02, 0.05, 0.001, 0.002}, and the $L_2$ regularization coefficient $β$ is chosen from {0.01, 0.005, 0.001, 0.0005}. The NO-HGNN model is configured with 2 layers, i.e., $l$=2, and the feature dimension is fixed at 32. Hyperparameters are carefully tuned to ensure optimal performance for each model.
c) The termination conditions are defined as follows: the training process halts if the iteration threshold reaches 300 or if no improvement in performance is observed for 10 consecutive iterations.

## B. Comparison Results

To assess the link prediction capabilities of the proposed NO-HGNN, we compare it against six baseline models. Table II presents the F1-Score and Accuracy achieved by all models on the test set. Based on the results, two key observations can be made:
a) Compared to the two static graph neural networks (M1 and M2), M7 demonstrates superior performance in DG learning for link prediction. For example, on D1 (Table II), M7 obtains an F1-Score of 0.8301, outperforming M1 and M2, which score 0.7641 and 0.7220, respectively. Consistent results are observed across other datasets. This improvement is largely due to the inability of M1 and M2 to capture the temporal and structural dynamics of evolving graphs, resulting in less effective representations.
b) Compared to the four dynamic GNN baselines (M3-M6), M7 obtains a notable performance boost. These models typically rely on normalized adjacency matrices or basic node feature similarities for message passing. As illustrated in Table II for dataset D1, M7 achieves an F1-Score of 0.8301 and an Accuracy of 0.8125, surpassing the performance of M3-M6. Similar improvements are observed across other datasets. This advantage stems from M7's ability to adopt neighbor overlap correlations in the message-passing process, enabling more precise DG learning and improving link prediction accuracy, an aspect overlooked by the baseline methods.

## V. CONCLUSION

In this study, a Neighborhood Overlap-Aware High-Order Graph Neural Network (NO-HGNN) is propsed for dynamic graph learning. By quantifying neighborhood overlap and integrating the resulting correlation scores into message passing, NO-HGNN learns more informative node embeddings. Experiments on dynamic graph benchmarks demonstrate improved link prediction accuracy. For future work, we plan to enhance NO-HGNN by:
a) evaluating its performance on a wide range of DG datasets to verify its generalizability across different domains;
b) conducting ablation studies to isolate the contributions of neighborhood overlap estimation and high-order message passing;
c) improving its scalability and computational efficiency for large-scale dynamic graphs by graph transformer.

## REFERENCES

[1] J. Chen, X. Luo, Y. Yuan, M. Shang, Z. Ming, Z. Xiong, "Performance of latent factor models with extended linear biases," *Knowledge-Based Systems*, vol. 123, pp. 128-136, 2017.
[2] T. N. Kipf and M. Welling, "Semi-supervised classification with graph convolutional networks," in *Proc. of the Int. Conf. on Learning Representations*, pp. 1-14, 2017.
[3] T. He, Y. S. Ong, and L. Bai, "Learning conjoint attentions for graph neural nets," in *Proc. of Conf. on Neural Information Processing Systems*, 2021.
[4] J. Chen, X. Luo, Y. Yuan, Z. Wang, "Enhancing graph convolutional networks with an efficient k-hop neighborhood approach," *Information Fusion*, vol. 124, 2025.
[5] X. Xu, M. Lin, Z. Xu, and X. Luo, "Attention-mechanism-based neural latent-factorization-of-tensors model," ACM Trans. Knowl. Discov. Data, pp. 1-27, 2025.

[6] Y. Yuan, X. Luo, M. Shang, and D. Wu, "A generalized and fast-converging non-negative latent factor model for predicting user preferences in recommender systems," in *Proc.of The ACM Web Conf.* pp. 498-507, 2020.
[7] F. Bi, T. He, Y. S. Ong, and X. Luo, "Graph linear convolution pooling for learning in incomplete high-dimensional data," *IEEE Trans. on Knowledge and Data Engineering*, vol. 37, no. 4, pp. 1838-1852, 2025.
[8] Y. Xia, M. Zhou, X. Luo, S. Pang and Q. Zhu, "A stochastic approach to analysis of energy-aware DVS-enabled cloud datacenters," *IEEE Transactions on Systems, Man, and Cybernetics: Systems*, vol. 45, no. 1, pp. 73-83, 2015.
[9] J. Chen, Y. Yuan, T. Ruan, J. Chen, and X. Luo, "Hyper-parameter-evolutionary latent factor analysis for high-dimensional and sparse data from recommender systems," *Neurocomputing*, 2020, 421: 316-328.
[10] X. Xu, T. Zhang, C. Xu, Z. Cui, and J. Yang, "Spatial-temporal tensor graph convolutional network for traffic speed prediction," *IEEE Trans. on Intelligent Transportation Systems*, vol. 24, no. 1, pp. 92-103, 2023.
[11] D. Wu, M. Shang, X. Luo, and Z. Wang, "An $L_1$-and-$L_2$-norm-oriented latent factor model for recommender systems," *IEEE Trans. on Neural Networks and Learning Systems*, vol. 33, no. 10, pp. 5775-5788, 2022.
[12] Y. Yuan, X. Luo, and M. Zhou, "Adaptive divergence-based non-negative latent factor analysis of high-dimensional and incomplete matrices from industrial applications," *IEEE Trans. on Emerging Topics in Computational Intelligence*, vol. 8, no.2, pp. 1209-1222. 2024.
[13] D. Wu, X. Luo, Y. He, and M. Zhou, "A prediction-sampling-based multilayer-structured latent factor model for accurate representation of high-dimensional and sparse data," *IEEE Trans. on Neural Networks and Learning Systems*, vol. 35, no. 3, pp. 3845-3858, 2024.
[14] X. Luo, H. Wu, Z. Wang, J. Wang, and D. Meng, "A Novel approach to large-scale dynamically weighted directed network representation," *IEEE Trans. on Pattern Analysis and Machine Intelligence*, vol. 44, no. 12, pp. 9756-9773, 2022.
[15] Y. Yuan, X. Luo, and M. Shang, "Effects of preprocessing and training biases in latent factor models for recommender systems," *Neurocomputing*, pp. 2019-2030, 2018.
[16] F. Bi, T. He, Y. Xie, and X. Luo, "Two-stream graph convolutional network-incorporated latent feature analysis," *IEEE Trans. on Services Computing*, vol. 16, no. 4, pp. 3027-3042, 2023.
[17] Z. Li, S. Li and X. Luo, "A novel machine learning system for industrial robot arm calibration," *IEEE Trans. on Circuits and Systems II: Express Briefs*, vol. 71, no. 4, pp. 2364-2368, 2024.
[18] X. Luo, H. Wu and Z. Li, "Neulft: A novel approach to nonlinear canonical polyadic decomposition on high-dimensional incomplete tensors," *IEEE Trans. on Knowledge and Data Engineering*, vol. 35, no. 6, pp. 6148-6166, 2023.
[19] F. Bi, X. Luo, B. Shen, H. Dong, and Z. Wang, "Proximal alternating-direction-method-of-multipliers-incorporated nonnegative latent factor analysis," *IEEE/CAA Journal of Automatica Sinica*, vol. 10, no. 6, pp. 1388-1406, 2023.
[20] Y. Yuan, S. Lu, and X. Luo, "A proportional integral controller-enhanced non-negative latent factor analysis model, *IEEE/CAA Journal of Automatica Sinica*," DOI: 10.1109/JAS.2024.125055, 2024.
[21] X. Luo, H. Wu, H. Yuan, and M. Zhou, "Temporal pattern-aware QoS prediction via biased non-negative latent factorization of tensors," *IEEE Trans. on Cybernetics*, vol. 50, no. 5, pp. 1798-1809, 2020.
[22] J. Li, X. Luo, Y. Yuan, and S. Gao, "A nonlinear PID-incorporated adaptive stochastic gradient descent algorithm for latent factor analysis," *IEEE Trans. on Automation Science and Engineering*, vol. 21, no. 3, pp. 3742-3756, 2024.
[23] T. He, Y. Liu, Y. S. Ong, X. Wu, and X. Luo, "Polarized message-passing in graph neural networks," *Artificial Intelligence*, vol. 331, p. 104129, 2024.
[24] X. Luo, Y. Yuan, S. Chen, N. Zeng, and Z. Wang, "Position-transitional particle swarm optimization-incorporated latent factor analysis," *IEEE Trans. on Knowledge and Data Engineering*, vol. 34, no. 8, pp. 3958-3970, 2022.
[25] M. Chen, L. Tao, J. Lou, and X. Luo, "Latent-Factorization-of-Tensors-Incorporated Battery Cycle Life Prediction," IEEE/CAA Journal of Automatica Sinica, vol. 12, no. 3, pp. 633-635, 2025.
[26] Y. Yuan, R. Wang, G. Yuan, and X. Luo, "An adaptive divergence-based non-negative latent factor model," *IEEE Trans. on Systems, Man, and Cybernetics: Systems*, vol. 53, no. 10, pp. 6475-6487, 2023.
[27] X. He, K. Deng, X. Wang, Y. Li, Y. Zhang, and M. Wang, "LightGCN: Simplifying and powering rgraph convolution network for recommendation", in *Proc. of the ACM SIGIR Conf. on Research and Development in Information Retrieval*, pp. 639-648, 2020.
[28] J. Li, Y. Yuan, T. Ruan, J. Chen, and X. Luo, "A proportional-integral-derivative-incorporated stochastic gradient descent-based latent factor analysis model," *Neurocomputing*, vol. 427, pp. 29-39, 2021.
[29] D. Wu, Y. He, X. Luo, and M. Zhou, "A Latent Factor Analysis-Based Approach to Online Sparse Streaming Feature Selection," IEEE Transactions on Systems, Man, and Cybernetics: Systems, vol. 52, no. 11, pp. 6744-6758, 2022.
[30] R. Xu, D. Wu, R. Wang and X. Luo, "A highly-accurate three-way decision-incorporated online sparse streaming features selection model," *IEEE Trans. on Systems, Man, and Cybernetics: Systems*, doi: 10.1109/TSMC.2025.3548648, 2025.
[31] Y. Yuan, Q. He, X. Luo, and M. Shang, "A multilayered-and-randomized latent factor model for high-dimensional and sparse matrices," *IEEE Trans. on Big Data*, vol. 8, no. 3, pp. 784-794, 2022.
[32] D. Wu, Z. Li, Z. Yu, Y. He, and X. Luo, "Robust Low-Rank Latent Feature Analysis for Spatiotemporal Signal Recovery," IEEE Transactions on Neural Networks and Learning Systems, vol. 36, no. 2, pp. 2829-2842, 2025.
[33] M. Shang, Y. Yuan, X. Luo, and M. Zhou, "An α-β-divergence-generalized recommender for highly accurate predictions of missing user preferences," *IEEE Trans. on Cybernetics*, vol. 52, no. 8, pp. 8006-8018, 2022.
[34] D. Wu, X. Luo, M. Shang, Y. He, G. Wang and M. Zhou, "A deep latent factor model for high-dimensional and sparse matrices in recommender systems," *IEEE Trans. on Systems, Man, and Cybernetics: Systems*, vol. 51, no. 7, pp. 4285-4296, 2021.
[35] X. Liao, K. Hoang and X. Luo, "Local Search-Based Anytime Algorithms for Continuous Distributed Constraint Optimization Problems," *IEEE/CAA Journal of Automatica Sinica*, vol. 12, no. 1, pp. 288-290, 2025.
[36] D. Wu, X. Luo, G. Wang, M. Shang, Y. Yuan and H. Yan, "A highly accurate framework for self-labeled semisupervised classification in industrial applications," *IEEE Trans. on Industrial Informatics*, vol. 14, no. 3, pp. 909-920, 2018.
[37] J. Li, Y. Yuan, and X. Luo, "Learning error refinement in stochastic gradient descent-based latent factor analysis via diversified pid controllers," *IEEE Trans. on Emerging Topics in Computational Intelligence*, doi: 10.1109/TETCI.2025.3547854.
[38] D. Wu, W. Sun, Y. He, Z. Chen, and X. Luo, "MKG-FENN: a multimodal knowledge graph fused end-to-end neural network for accurate drug-drug interaction prediction," in *Proc. of the AAAI Conf. on Artificial Intelligence*, pp. 10216-10224, 2024.
[39] F. Manessi, A. Rozza, and M. Manzo, "Dynamic graph convolutional networks," *Pattern Recognition*, vol. 97, 2020.
[40] A. Pareja, G. Domeniconi, J. Chen, T. Ma, T. Suzumura, H. Kanezashi, T. Kaler, T. B. Schardl, C. E. Leiserson, "EvolveGCN: Evolving graph convolutional networks for dynamic graphs," in *Proc. of the AAAI Conf. on Artificial Intelligence*, pp. 5363-5370, 2020.
[41] Y. Yuan, M. Shang, and X. Luo, "Temporal web service QoS prediction via Kalman filter-incorporated dynamic latent factor analysis," in *Proc of European Conf. on Artificial Intelligence*, pp. 561-568, 2020.
[42] T. Chen, W. Yang, S. Li, and X. Luo, "An Adaptive p-Norms-Based Kinematic Calibration Model for Industrial Robot Positioning Accuracy Promotion," *IEEE Transactions on Systems, Man, and Cybernetics: Systems*, vol. 55, no. 4, pp. 2937-2949, 2025.
[43] R. Xu, D. Wu, and X. Luo, "Online sparse streaming feature selection via decision risk," 2023 IEEE International Conference on Systems, Man, and Cybernetics, pp. 4190-4195, 2023.
[44] J. Chen, K. Liu, X. Luo, Y. Yuan, K. Sedraoui, Y. Al-Turki, and M. Zhou, "A State-migration particle swarm optimizer for adaptive latent factor analysis of high-dimensional and incomplete data," *IEEE/CAA Journal of Automatica Sinica*, vol. 11, no. 11, pp. 2220-2235, 2024.

[45] F. Bi, T. He, and X. Luo, "A fast nonnegative autoencoder-based approach to latent feature analysis on high-dimensional and incomplete data," *IEEE Trans. on Services Computing*, vol. 17, no. 3, pp. 733-746, 2024.
[46] Y. Yuan, X. Luo, M. Shang, and Z. Wang, "A Kalman-filter-incorporated latent factor analysis model for temporally dynamic sparse data," *IEEE Trans. on Cybernetics*, vol. 53, no. 9, pp. 5788-5801, 2023.
[47] A. Sankar, Y. Wu, L. Gou, W. Zhang, and H. Yang, "DySAT: deep neural representation learning on dynamic graphs via self-attention networks," in *Proc. of the Int. Conf. on Web Search and Data Mining*, pp. 519-527, 2020.
[48] X. Luo, J. Chen, Y. Yuan and Z. Wang, "Pseudo gradient-adjusted particle swarm optimization for accurate adaptive latent factor analysis," *IEEE Trans. on Systems, Man, and Cybernetics: Systems*, vol. 54, no. 4, pp. 2213-2226, 2024.
[49] P. Tang and X. Luo, "Neural tucker factorization," *IEEE/CAA Journal of Automatica Sinica*, vol. 12, no. 2, pp. 475-477, 2025.
[50] X. Luo, Y. Yuan, M. Zhou, Z. Liu, and M. Shang, "Non-negative latent factor model based on $\beta$-divergence for recommender systems," *IEEE Trans. Systems, Man, and Cybernetics: Systems*, vol. 51, no. 8, pp. 4612-4623, 2021.
[51] D. Wu, Y. Hu, K. Liu, J. Li, X. Wang, S. Deng, N. Zheng, and X. Luo, "An outlier-resilient autoencoder for representing high-dimensional and incomplete data," *IEEE Trans. on Emerging Topics in Computational Intelligence*, doi: 10.1109/TETCI.2024.3437370, 2024.
[52] L. Wang, K. Liu, and Y. Yuan, "GT-A$^2$T: Graph tensor alliance attention network," *IEEE/CAA Journal of Automatica Sinica*, DOI: 10.1109/JAS.2024.124863, 2024.
[53] Z. Liu, X. Luo and M. Zhou, "Symmetry-constrained Non-negative Matrix Factorization Approach for Highly-Accurate Community Detection," *2021 IEEE 17th International Conference on Automation Science and Engineering*, 2021, pp. 1521-1526.
[54] D. Wu, M. Shang, X. Luo, J. Xu, H. Yan, W. Deng, G. Wang, "Self-training semi-supervised classification based on density peaks of data," *Neurocomputing*, Vol. 275, pp.180-191, 2018.
[55] J. Chen, Y. Yuan, and X. Luo, "SDGNN: symmetry-preserving dual-stream graph neural networks," *IEEE/CAA Journal of Automatica Sinica*, no. 11, vol. 7, 1717-1719, 2024.
[56] Z. Liu, X. Luo, S. Li and M. Shang, "Accelerated Non-negative Latent Factor Analysis on High-Dimensional and Sparse Matrices via Generalized Momentum Method," *2018 IEEE International Conference on Systems, Man, and Cybernetics*, 2018, 3051-3056.
[57] X. Luo, Z. Ming, Z. You, S. Li, Y. Xia, H. Leung, "Improving network topology-based protein interactome mapping via collaborative filtering," *Knowledge-Based Systems*, vol. 90, pp. 23-32, 2015.
[58] R. Xu, D. Wu and X. Luo, "Recursion-and-Fuzziness Reinforced Online Sparse Streaming Feature Selection," *IEEE Trans. on Fuzzy Systems*, doi: 10.1109/TFUZZ.2025.3569272.
[59] X. Luo, M. Zhou, S. Li, D. Wu, Z. Liu and M. Shang, "Algorithms of unconstrained non-negative latent factor analysis for recommender systems," *IEEE Transactions on Big Data*, vol. 7, no. 1, pp. 227-240, 2021.
[60] W. Qin, X. Luo and M. Zhou, "Adaptively-accelerated parallel stochastic gradient descent for high-dimensional and incomplete data representation learning," *IEEE Transactions on Big Data*, vol. 10, no. 1, pp. 92-107, 2024.
[61] Z. Liu, G. Yuan, and X. Luo, "Symmetry and nonnegativity-constrained matrix factorization for community detection," *IEEE/CAA Journal of Automatica Sinica*, vol. 9, no. 9, pp. 1691-1693, 2022.
[62] M. Shang, X. Luo, Z. Liu, J. Chen, Y. Yuan and M. Zhou, "Randomized latent factor model for high-dimensional and sparse matrices from industrial applications," *IEEE/CAA Journal of Automatica Sinica*, vol. 6, no. 1, pp. 131-141, 2019.
[63] Y. Yuan, Y. Wang, and X. Luo, "A node-collaboration-informed graph convolutional network for highly accurate representation to undirected weighted graph," *IEEE Trans. on Neural Networks and Learning Systems*, DOI: 10.1109/TNNLS.2024.3514652, 2024.
[64] Z. Liu, X. Luo and M. Zhou, "Symmetry and graph bi-regularized non-negative matrix factorization for precise community detection," *IEEE Transactions on Automation Science and Engineering*, vol. 21, no. 2, pp. 1406-1420, 2024.
[65] W. Yang, S. Li and X. Luo, "Vibration Control using A Robust Input Shaper via Extended Kalman Filter-Incorporated Residual Neural Network," *IEEE International Conference on Systems, Man, and Cybernetics*, 2024, pp. 3980-3986.
[66] F. Bi, T. He and X. Luo, "A two-stream light graph convolution network-based latent factor model for accurate cloud service qos estimation," *IEEE International Conference on Data Mining*, 2022, pp. 855-860.
[67] Z. Luo, X. Jin, Y. Luo, Q. Zhou, and X. Luo, "Analysis of students' positive emotion and smile intensity using sequence-relative key-frame labeling and deep-asymmetric convolutional neural network," *IEEE/CAA J. Autom. Sinica*, vol. 12, no. 4, pp. 806–820, 2025.
[68] Z. Liu, X. Luo, Z. Wang, X. Liu, "Constraint-induced symmetric nonnegative matrix factorization for accurate community detection," *Information Fusion*, vol. 89, pp. 588-602, 2023.